\documentclass[conference,a4paper]{IEEEtran}

\IEEEoverridecommandlockouts 

\usepackage[utf8]{inputenc}
\usepackage{multirow}
\usepackage{rotating}
\usepackage[T1]{fontenc}
\usepackage{graphicx}
\usepackage[table]{xcolor}
\usepackage{fancyhdr}
\usepackage{amsmath}
\usepackage{float}

\title{\LARGE \bf
CheX-Nomaly: Segmenting Lung Abnormalities from Chest Radiographs using Machine Learning}

\author{Sanskriti Singh$^{1}$}

\fancypagestyle{FirstPage}{}

\begin{document}

\maketitle
\thispagestyle{empty}
\pagestyle{empty}

\begin{abstract}

The global challenge in chest radiograph X-ray (CXR) abnormalities often being misdiagnosed is primarily associated with perceptual errors – where healthcare providers struggle to accurately identify the location of abnormalities – rather than misclassification errors. We currently address this problem through disease specific segmentation models. Unfortunately these models cannot be released in the field due to their lack of generalizability across all thoracic diseases. A binary model tends to perform poorly when it encounters a disease that isn't represented in the dataset. I present CheX-nomaly: a binary localization U-net model which leverages transfer learning techniques with the incorporation of an innovative contrastive learning approach. Trained on the VinDr-CXR dataset, which encompasses 14 distinct diseases in addition to ’no finding’ cases, my model achieves generalizability across these 14 diseases and others it has not seen before. I show that I can significantly improve the generalizability of an abnormality localization model by incorporating a contrastive learning method and dissociating the bounding boxes with its disease class. I also introduce a new loss technique to apply to enhance the U-nets performance on bounding box segmentation. By introducing CheX-nomaly, I offer a promising solution to enhance the precision of chest disease diagnosis, with a specific focus on reducing the significant number of perceptual errors in healthcare.

\end{abstract}

\section{INTRODUCTION}
\thispagestyle{FirstPage}

In spite of the abundance of diagnostic models available, their implementation in clinical settings confronts substantial challenges. A prominent issue pertains to the peril of misdiagnosis, carrying potentially severe consequences for patient outcomes [4]. Moreover, the application of artificial intelligence in healthcare transcends singular pathologies or medical conditions. Efforts have been redirected towards crafting models capable of simultaneous detection and segmentation of multiple diseases to ameliorate these limitations and enhance the practicality of diagnostic models. However, the performance of these multi-disease models falls short when compared to their binary counterparts, necessitating further research to ensure clinical effectiveness and reliability [5].

\begin{figure}[H]
    \centering
    \includegraphics[width=0.3\textwidth]{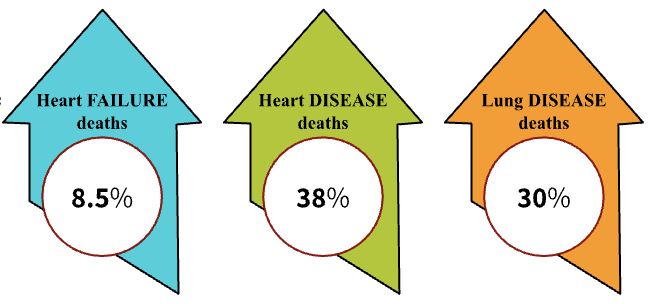} 
    \caption{Thoracic disease is a prominent cause of mortality in the United States and a significant contributor to diagnostic errors.}
\end{figure}

Thoracic diseases present a substantial global healthcare burden, accounting for over 42.6\% of all deaths in the United States and standing as the primary cause of worldwide mortality [13]. As depicted in Fig. 1, the mortality rates of heart failure, heart disease, and lung disease within the U.S. are highlighted. Although chest radiographs serve as the gold standard for diagnosing thoracic conditions, leading to the development of numerous models, their limited generalization across various disease types found in CXRs hinders the creation of deployable models, despite extensive research in this domain.

\begin{figure}
    \centering
    \includegraphics[width=0.4\textwidth]{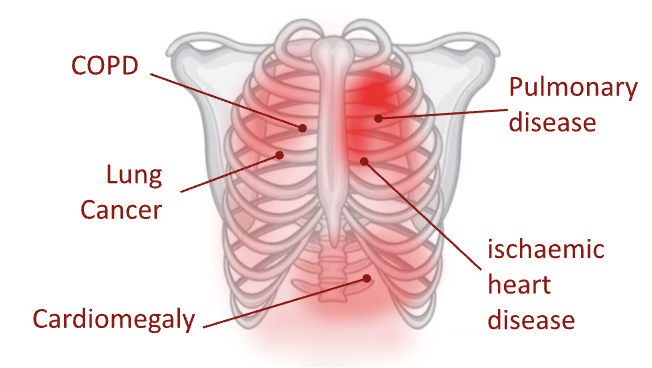} 
    \caption{The red shows areas in which more diseases have been overlooked. Darker colors refer to more perceptual errors [citation here]}
\end{figure}

The Radiological Society of North America (RSNA) has actively distributed datasets to advance localization models, aiming to facilitate the development of robust models [10]. Prominent models in lung disease diagnosis encompass VDSNet from Bangladesh University of Engineering and Technology [1], pneumonia detection models employing the RSNA dataset, and lung segmentation models utilizing the Japanese Society of Radiological Technology (JSRT) dataset [6].

Perceptual errors, reported by the National Institutes of Health (NIH) as constituting 80\% of all misdiagnoses, occur when radiologists encounter challenges in precisely localizing anomalies within CXRs, despite their proficiency in identifying abnormalities or diseases. Fig. 2 illustrates common areas of perceptual errors leading to misdiagnosis or suboptimal patient care. In contrast, misclassification errors, accounting for a smaller share of diagnostic mistakes (13\% of cases), arise when radiologists struggle to differentiate between the correct abnormality and another disease. Given the lower prevalence of misclassification errors, the primary focus in enhancing diagnostic models should center on addressing perceptual errors.

Notably, diagnostic models designed to address perceptual errors must prioritize generalization. Disease-specific models excel in recognizing particular diseases but often falter in accurately localizing anomalies not explicitly linked to those diseases, impeding the broader category of misdiagnosis: perceptual errors. Effectively mitigating misdiagnoses necessitates a shift towards identifying and precisely localizing any abnormality within the images, rather than solely focusing on abnormalities associated with specific diseases. This strategy aligns with the primary objective of diminishing perceptual errors throughout the diagnostic process.

\begin{figure}[H]
    \centering
    \includegraphics[width=0.45\textwidth]{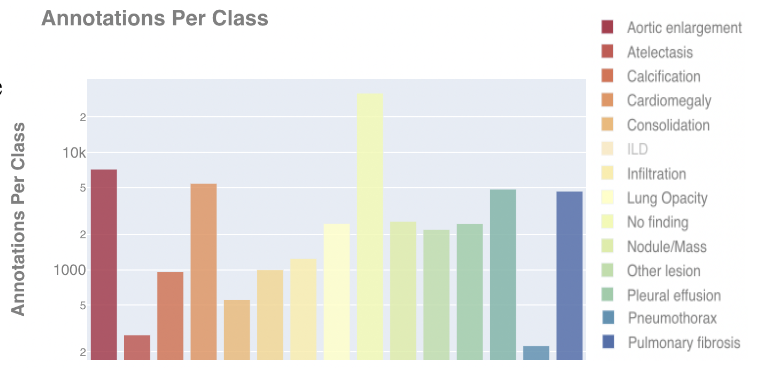} 
    \caption{Distribution of annotations (one bounding box) for each disease. Each image can have more than 1 bounding box or disease.}
\end{figure}

Introducing "CheX-nomaly," a tool specifically crafted for precise abnormality localization within chest radiograph X-rays (CXRs), irrespective of disease classification. Harnessing diverse machine learning methodologies such as the Attention U-net architecture, contrastive learning technique, and knowledge distillation, CheX-nomaly surpasses existing multi-disease and specialized binary models. The development and deployment of CheX-nomaly show promise in enabling precise abnormality localization, thus reducing the incidence of misdiagnosis. This advancement holds the potential to establish new standards in the realm of respiratory disease diagnosis.

\section{Related Work}

Current research frequently tackles both perceptual and classification errors in chest radiography; however, the persisting challenge of limited generalizability remains. These models, akin to their classification-focused counterparts, often learn disease-specific features and struggle when encountering previously unseen diseases.

Significant investigation has utilized the VinDr-CXR dataset [14], also employed in this project, crafted to cater to both classification and localization tasks. The most successful model on this dataset achieves a mean Average Precision (mAP) of 0.314, following the standard PASCAL VOC 2010 mAP metric at Intersection over Union (IoU) greater than 0.4 [8]. Competing closely, the second and third-place models achieve mAP values of 0.307 [15] and 0.305 [11]), respectively.

Simultaneously, models exclusively trained for disease classification have gained prominence, especially utilizing datasets such as RSNA and NIH for demanding diagnostic classification models. For instance, CheX-Net, developed by Stanford University, demonstrates an impressive F1 score of 0.435 across 14 distinct thoracic diseases [7]. Additionally, my prior work on pneumonia with PneumoXttention displays an F1 score of 0.82 [12].

CheX-Nomaly rectifies the lack of research dedicated to the generalization of these models by emphasizing the localization of abnormalities, irrespective of the specific disease. This strategy signifies a significant advancement in mitigating perceptual errors and augmenting the generalizability of anomaly localization, contributing markedly to the progression of respiratory disease diagnosis.

\section{Data}

\begin{figure}[H]
    \centering
    \includegraphics[width=0.48\textwidth]{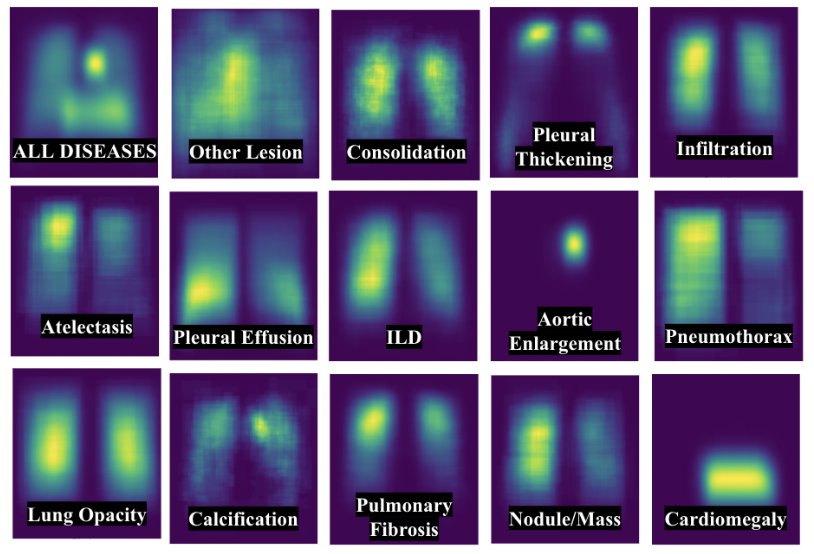} 
    \caption{The heatmpas illustrate the locations of corresponding diseases, with brighter shades of green indiciating a higher concerntration of abnromaltieis in a specific area across the VinDr-CXR dataset.}
\end{figure}

The project leveraged the VinDr-CXR dataset, an openly accessible repository of chest X-rays featuring annotations provided by radiologists [source: Kaggle VinBigData Chest X-ray Abnormalities Detection]. This dataset encompasses 18,000 postero-anterior (PA) chest X-rays in DICOM format, each examined and annotated by multiple radiologists, resulting in up to 58 distinct annotations stored as bounding boxes (Fig. 3). Each scan was designated a label corresponding to one of 14 respiratory diseases or categorized as 'no finding' (Fig. 4). 

The VinDr-CXR dataset was meticulously curated from over 100,000 raw DICOM images retrospectively obtained from two prominent medical institutions in Vietnam: Hospital 108 and the Hanoi Medical University Hospital. Seventeen highly experienced radiologists, each with a minimum of 8 years of professional experience, conducted the annotations. These experts identified 14 critical findings (local labels) and classified 6 distinct diagnoses (global labels). Each critical finding was pinpointed using a bounding box. Specifically, these local and global labels align with the 'Findings' and 'Impressions' sections, respectively, commonly present in standard radiology reports. 'Findings' encompass detailed observations or abnormalities identified in the medical images, while the 'Impressions' offer a succinct summary or diagnosis based on those findings. However, for this project's scope, focus was solely placed on the 'Findings' section.

\begin{figure}[H]
    \centering
    \includegraphics[width=0.48\textwidth]{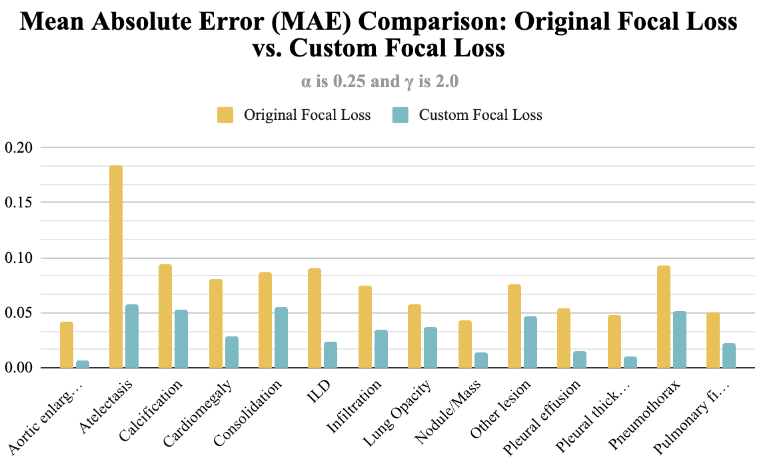} 
    \caption{The mean absolute error with the customized loss function is significantly reduced.}
\end{figure}

The dataset underwent diverse divisions to accommodate different methodologies. Initially, the entire dataset was split in an 80-10-10 ratio for training, validation, and testing to ensure uniformity across all models. Subsequently, the 80\% training data underwent further manipulation according to the specific methodology adopted. For example, specialized models utilized datasets with a 50-50 split between disease and no-disease data, with each model having its unique dataset. Conversely, the supermodel was not subjected to additional data manipulation and was trained on the entire 80\% available data. To optimize this supermodel through transfer learning with a siamese network, a new dataset comprising two randomly selected scans and two classes (similar and dissimilar) was created from the training set.

The test set, constituting 10\% of the entire dataset, remained consistent across all three models to establish a foundational comparison base. To assess the model's generalizability, tests were conducted on diseases to which the models had not been previously exposed, such as pneumonia. For this purpose, pneumonia scans from the RSNA dataset released by Kaggle in 2018 were utilized [9]. This dataset contained 26,684 frontal-view X-ray images encompassing both pneumonia cases and cases without pneumonia. Notably, the subset labeled as "without pneumonia" did not necessarily indicate the absence of other diseases, as there were no specific annotations for these other diseases. Consequently, only scans with pneumonia were used for evaluation.

To maintain the ratio of positive to negative images in the original test set (4:10), healthy patient data was obtained from the CANDID-PTX private dataset [2] and the University of Montreal COVID Chest X-ray Dataset [3]. The majority of the testing primarily utilized the Pneumonia dataset, as it provided a broader selection of pre-trained models, simplifying comparative analyses.

\section{Methodology}
\subsection{Problem Formation}
`
The thoracic disease localization task addressed by CheX-Nomaly is a binary object detection problem. In this context, the input is a 512x512 grayscale frontal-view chest X-ray image \(X\), and the output is a 512x512 binary mask \(y\) where \(y \in \{0,1\}\), signifying the absence or presence of an abnormality.

For each training example, a custom loss function was employed using the sigmoid focal cross-entropy loss from TensorFlow Addons (Tfa.losses.sigmoidfocalcrossentropy). The loss function is defined as:

\[y' = \text{transform}(y_{\text{pred}})\]
\[L(y_{\text{true}}, y') = [-\alpha * (1 - y')^{\gamma} * \log(y') * y_{\text{true}}] + \]
\[[-(\alpha-1) * (1 - y')^{\gamma} * \log(y') * (1-y_{\text{true}})]\]

where:
\begin{itemize}
    \item \(y_{\text{true}}\) is the true binary label (0 for the negative class, 1 for the positive class).
    \item \(y_{\text{pred}}\) is the predicted probability of the example being in the positive class.
    \item \(y'\) is the predicted probability of the example being in the positive class after transformation.
    \item \(\alpha\) is the balancing factor controlling the weight given to each class.
    \item \(\gamma\) is the modulating factor amplifying the loss for challenging cases.
\end{itemize}

\begin{figure}[H]
    \centering
    \includegraphics[width=0.48\textwidth]{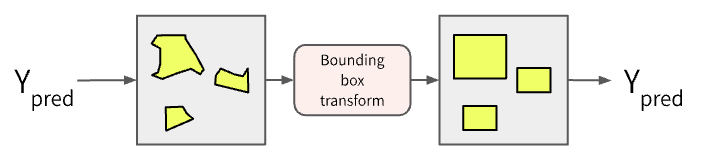} 
    \caption{Transformation of \(y_{\text{pred}}\) into more rigid rectangular shapes.}
\end{figure}

The hyperparameters \(\alpha\) and \(\gamma\) play crucial roles in this loss function. In experiments with a specialized model, it was found that setting \(\alpha = 0.25\) and \(\gamma = 2.0\) optimized the model's performance, aligning with default values in the code.

This tailored loss function introduces an adjustment mechanism to the model's predictions (\(y_{\text{pred}}\)). The purpose is to refine pixel-wise prediction masks, aligning them with the final binary masks representing abnormalities as bounding boxes (refer to Fig. 6). This adjustment enables the model to concentrate on abnormalities, which might have irregular shapes but can be accurately approximated by bounding boxes. This adaptation prevents the model from explicitly learning the shape of bounding boxes. Fig. 5 validates that the predicted rectangles closely correspond to actual disease patterns. This approach provides greater flexibility in representing shapes, deviating from rigid adherence to rectangular forms.

\subsection{Model Architecture and Training}
\subsection{\textbf{Specialized Models}}

The specialized models system is engineered to address both perceptual and classification errors within a computer vision framework. This framework operates by training a series of specialized models, each specifically tailored to identify a particular disease. These models are collectively applied to medical images to generate predictive masks. In a clinical setting, radiologists would integrate these predictions into their diagnostic workflow. However, a notable challenge emerges when these models encounter diseases beyond their individual expertise, potentially resulting in unpredictable final masks.

\begin{figure}[H]
    \centering
    \includegraphics[width=0.48\textwidth]{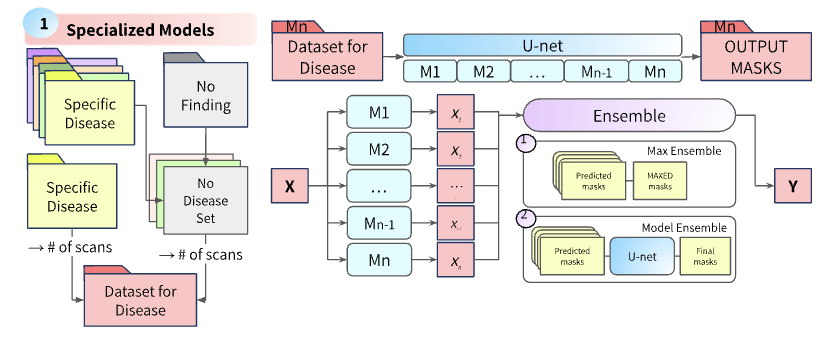} 
    \caption{The specialized models have U-net backbone and are assembled in one of 2 ways for final mask.}
\end{figure}

To empirically evaluate this approach, an experiment was conducted using the VinDr-CXR dataset. Fourteen specialized models were meticulously developed, all based on a U-net architecture. Each model was meticulously trained to focus on one of the 14 distinct diseases within the dataset. The training data for each specialized model included images paired with corresponding masks and a random selection of other scans, ensuring a balanced 50-50 ratio, as visually depicted in the left section of Fig. 7.

Following the training of all 14 specialized models, a new dataset was curated, encompassing 14 predicted masks and a unified output mask for each image in the complete dataset. These 14 predicted masks were then consolidated using one of two methods, as demonstrated in Fig. 7: 'Max Ensemble' and a 'Model Ensemble'.

\subsection{\textbf{Supermodels}}

In the pursuit of creating a more generalized model capable of handling a broad spectrum of diseases, an ideal scenario involves an infinitely expansive dataset that covers a comprehensive range of disease characteristics. Such a dataset would compel the model to prioritize generalized abnormality features over specific disease-related traits. Our "supermodel" approach aimed to feed all available data simultaneously to a single model without disease classification, as illustrated in Fig. 8. The primary objective was the identification of abnormalities in chest radiographs (CXRs) without prior knowledge of specific diseases. This strategy aimed to emancipate the model from the constraints of disease-specific detection, enabling it to focus on normal chest anatomy features rather than disease-specific characteristics.

\begin{figure}[H]
    \centering
    \includegraphics[width=0.35\textwidth]{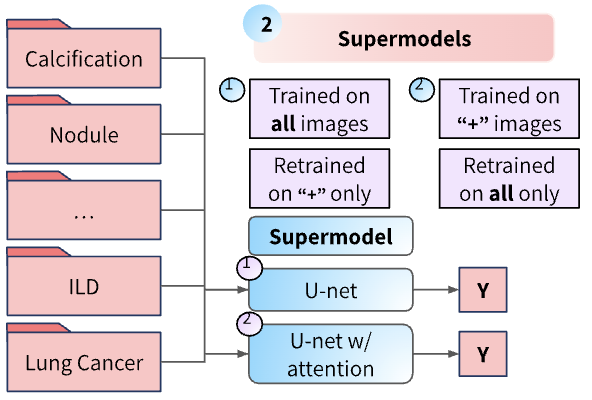} 
    \caption{The U-net supermodel took all the data as input. Due to data imbalance 2 contrasting methods of training was devised to force model to put emphasis on disease prediction. Two supermodels were tested w and w/o attention.}
\end{figure}

However, limitations imposed by finite data and extensive training introduced the possibility that the model might unintentionally learn disease-specific features associated with the 14 diseases under consideration. This circumstance could potentially hinder its ability to generalize across other diseases. Upon testing the model on the "other" dataset, it demonstrated a moderate capacity to detect some abnormalities. However, it did not match the performance levels achieved by specialized models explicitly trained for those diseases. Consequently, efforts were made to optimize the model's performance by integrating transformer-based attention blocks. These attention blocks were introduced to augment the model's focus on generalized disease features.

Upon analysis, it became apparent that the incorporation of attention blocks led to an improvement in the model's ability to identify abnormalities related to the 14 diseases it had been trained on. However, the model's performance exhibited a decline when confronted with previously unencountered diseases, indicating a potential inadvertent emphasis on more disease-specific features. This specialization affected the model's adaptability to novel conditions, necessitating a new approach to optimize this supermodel.

\subsection{\textbf{CheX-Nomaly}}

To augment the generalizability of our supermodel across a diverse spectrum of diseases, both known and unknown within its dataset, a novel contrastive learning-based approach was developed. This approach focused on the model's ability to discern between disease-specific attributes and general abnormalities, emphasizing the capacity to identify deviations from a healthy patient's chest X-ray (CXR). Initially, attempts to create a diverse dataset aimed to encourage the model to perceive general abnormalities as a unified class. However, during optimization, the model primarily focused on the 14 specific diseases.

\begin{figure}[H]
    \centering
    \includegraphics[width=0.48\textwidth]{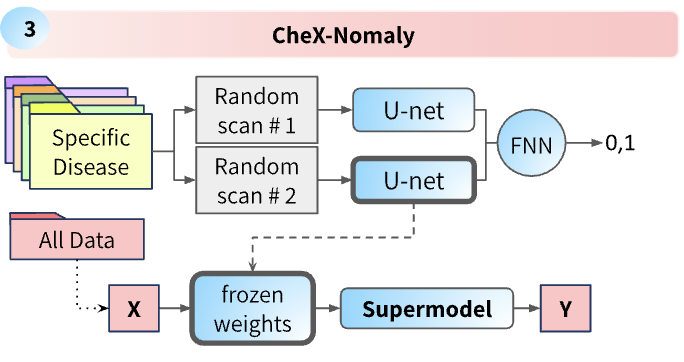} 
    \caption{CheX-Nomaly employs a transfer learning approach from a Siamese contrastive learning model and integrates elements from the original supermodel. In this context, (P,N) and (N,P) belong to class 0, while (P,P) belongs to class 1.}
\end{figure}

To tackle this issue, a novel approach was devised. Another model was trained to categorize pairs of scans based on their content: 0 for dissimilar data points and 1 for similar points. This auxiliary model was designed to learn similarities between different abnormalities and the contrast between healthy and unhealthy patients. The intention was to transfer this auxiliary model's knowledge to the supermodel by freezing its weights.

However, a challenge arose as the auxiliary model accepted two inputs, whereas the supermodel was designed for only one. To address this, a Siamese network architecture was introduced to connect the auxiliary model with the supermodel. This Siamese network was trained to distinguish similarities among different disease-related CXRs while simultaneously identifying dissimilarities between diseases and normal CXRs, employing a contrastive loss function:

\[L(x_1,x_2,y) = \frac{1}{2}* y * D^2 + \frac{1}{2} * (1-y) * \max(0,\phi - D)^2\]

Where:
\begin{itemize}
    \item \(x_1\) and \(x_2\) are the feature vectors of the input images.
    \item \(y\) is the binary label indicating similarity (1) or dissimilarity (0).
    \item \(D\) is the Euclidean distance between the feature vectors.
    \item \(\phi\) is the hyperparameter controlling the separation between similar and dissimilar pairs.
\end{itemize}

The Siamese network architecture comprises two identical base models, processing 512x512 input images through convolutional and dense layers. Their outputs are concatenated, followed by a dense layer and sigmoid activation function, yielding a classification output with two classes:

\textbf{Class 0} - Disease and normal (Dissimilar)

\textbf{Class 1} - Disease and disease (Similar)

The Siamese network was adapted for image classification, designed to measure the dissimilarity between pairs of input data.

In the final model architecture, the pre-trained Siamese network was integrated and configured to freeze the left-side layers, serving as a feature extractor. Extracted features were reshaped and up-sampled to align with the U-Net architecture (Fig. 9).

Upon evaluation using a distinct dataset, the new supermodel displayed enhanced generalization compared to the original. While the original supermodel excelled on the 14 trained diseases, it struggled on other diseases. Conversely, the new supermodel showcased improved generalization, signifying a significant advancement in medical image analysis (Table 1).

\subsection{\textbf{Model Architecture and Training}}

\begin{figure}
    \centering
    \includegraphics[width=0.48\textwidth]{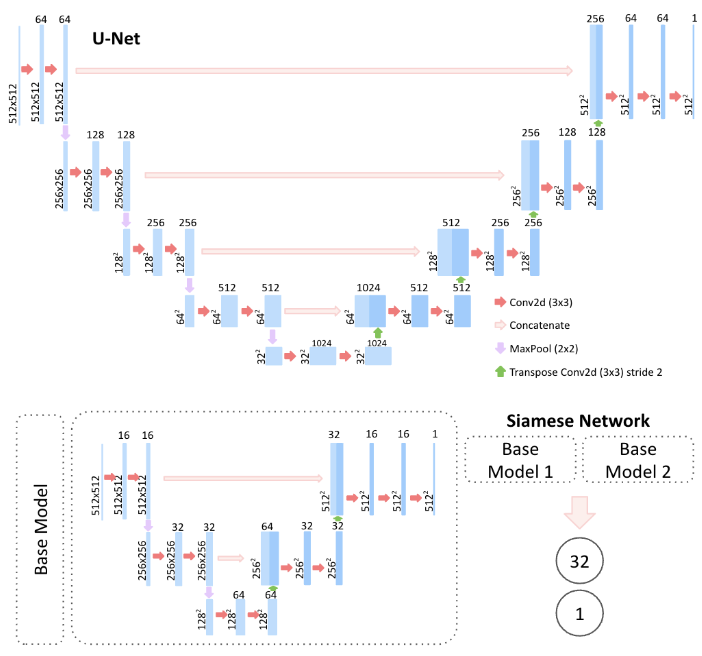} 
    \caption{Base model architectures of U-Net and siamese network}
\end{figure}

Fig. 10 presents a uniform training methodology adopted across all models, despite the inherent disparity in the methods used. Each model within these methodologies was trained using an identical U-Net architecture, coupled with similar parameter configurations fine-tuned until the validation loss plateaued. This standardized training protocol nullifies the influence of architectural discrepancies or optimization techniques. Thus, the singular distinguishing factor among the models remains confined to a single variable.

\begin{table*}
\centering
\caption{Statistical Results for All Models Across Both Datasets}
\label{tab:my-table7}
\begin{tabular}{|l|l|l|l|l|l|l|l|}
\hline
\multirow{2}{*}{} & \multirow{2}{*}{} & \multicolumn{2}{c|}{\textbf{Specialized Models}} & \multicolumn{2}{c|}{\textbf{Super Model}} & \multicolumn{2}{c|}{\textbf{Contrastive Learning}} \\ \cline{3-8} 
 &  & \textbf{Model Ensemble} & \textbf{U-net} & \textbf{Attention} & \textbf{Difference} & \textbf{U-net VinDr-CXR} & \textbf{Super Model} \\ \hline
\multirow{3}{*}{\textbf{Difference}} & \textbf{MAE} & 0.21 & 0.09 & 0.009 & 0.008 & 0.001 & 0.0109 \\ 
 & \textbf{F1 score (all)} & 0 & 0.11 & 0.23 & 0.23 & 0 & 0.58 \\ 
 & \textbf{Mean IOU} & 0.0102 & 0.0122 & 0.6118 & 0.6504 & -0.0386 & 0.7487 \\ \hline
\multirow{3}{*}{\textbf{Other Disease (200)}} & \textbf{MAE} & 0.408 & 0.37 & 0.204 & 0.272 & -0.068 & 0.131 \\ 
 & \textbf{F1 score (+)} & 1 & 0.96 & 0.96 & 0.96 & 0 & 0.98 \\ 
 & \textbf{Mean IOU} & 0.0091 & 0.0101 & 0.2119 & 0.1999 & 0.012 & 0.2801 \\ \hline
\end{tabular}
\end{table*}

The U-Net architecture, characterized by Conv2D layers employing 3x3 filters, concatenation operations, max-pooling, and transpose Conv layers, underwent slight adjustments contingent upon the specific methodology employed. For instance, the attention supermodel incorporated additional attention blocks at each skip connection, amplifying its functional capacity. Conversely, the Siamese network adapted a modified U-Net architecture to accommodate the memory constraints of the computer.

Initial experiments involved exploring a CNN with a smaller final feature map size before settling on the U-Net design. However, this alternative approach proved ineffective, as the resulting logits were insufficient to process a sizable input (512x512), impeding the model's ability to extract meaningful information from a mere 8 numbers.

\section{Results}

\subsection{\textbf{Specialized Models}}

The Max Ensemble method, upon evaluation on the VinDr-CXR test set, produced a Mean Absolute Error (MAE) of 0.21 and a Mean Intersection over Union (IoU) of 0.0102. Ideal models typically exhibit a MAE of 0 and an F1 score of 1. Detailed statistical analysis, including MAE and F1 scores for each disease within the test set (refer to Table 2 and Appendix A), demonstrated remarkable performance of specialized models across most diseases. Notably, the Nodule/Mass disease model presents an outlier due to limited data availability and a small bounding box-to-image ratio. However, the ensemble method's performance fell short of this near perfection due to its reliance on prior knowledge of specific diseases, making it redundant in scenarios where prior suspicion already exists. Visual representation of this method's performance in Fig. 11, column 3 (specialized models), indicates a lack of coherence in predictions, dispersed across the image without clear localization. While some bounding boxes approach perfection, predictions from unfamiliar specialized models result in an inaccurate final mask.

\begin{table}[H]
\centering
\caption{Mean Absolute Error and F1 Score for Each Specialized Model on Their Respective Disease Test Sets}
\label{tab:my-table8}
\begin{tabular}{|l|l|l|}
\hline
\textbf{Category} & \textbf{MAE} & \textbf{F1 score} \\ \hline
Aortic Enlargement & 0.0067 & 0.83 \\
Atelectasis & 0.058 & 0.83 \\
Calcification & 0.053 & 0.72 \\
Cardiomegaly & 0.0284 & 0.7 \\
Consolidation & 0.0554 & 0.78 \\
ILD & 0.0243 & 0.63 \\
Infiltration & 0.0341 & 0.72 \\
Lung Opacity & 0.0366 & 0.75 \\
Nodule/Mass & 0.0134 & 0 \\
Other lesion & 0.0469 & 0.64 \\
Pleural effusion & 0.0149 & 0.81 \\
Pleural thickening & 0.0099 & 0.71 \\
Pneumothorax & 0.0522 & 0.53 \\
Pulmonary fibrosis & 0.0225 & 0.76 \\ \hline
\end{tabular}
\end{table}

The second ensemble method, known as the Model Ensemble, achieved a mean absolute error of 0.09 and a mean intersection over union of 0.0122. However, this low MAE does not necessarily reflect improved prediction accuracy, but rather a tendency of the model to predict very little, resulting in a low MAE with a similar IoU. As seen in Fig. 11, row 4 (Normal CXR), there is a noticeable absence of true values among its predictions. Even if this method were to address these challenges, its generalization capabilities remain notably constrained, requiring retraining for each new abnormality. It is limited by a fixed number of detectable abnormalities, offering no assurance of seamless generalization to other diseases. For a detailed summary of both ensemble methods' performance, including their evaluation on the VinDr-CXR test set and an unseen dataset maintaining a consistent ratio of positive to negative images, refer to Table 1.

\subsection{\textbf{Supermodels}}

\begin{table}[H]
\centering
\caption{Comparison of the Top-Performing Models on the Pneumonia Dataset with State-of-the-Art Models}
\label{tab:mean-iou9}
\begin{tabular}{|l|l|}
\hline
\textbf{Models} & \textbf{Mean IOU} \\ \hline
CheX-Nomaly & 0.2801 \\
Supermodel & 0.2119 \\
Darapaneni1 et al & 0.32 \\
Tudouni et al & 0.26785 \\
IMVY [ods.ai] & 0.25595 \\
Save the alveoli! & 0.24351 \\ \hline
\end{tabular}
\end{table}

The initial supermodel showcased outstanding performance on the VinDr-CXR base test dataset, yielding a mean absolute error (MAE) of 0.009. After incorporating transformative attention blocks, a marginal reduction in MAE was achieved, reaching a value of 0.008. Both MAE values demonstrated substantial improvements compared to the specialized models, initially indicating comparable performance on the "Other dataset." However, the original supermodel achieved a MAE of 0.204 on this dataset, while the attention-enhanced supermodel registered a higher MAE of 0.212. A comprehensive model comparison across both datasets is provided in Table 1.

The numerical analysis overwhelmingly favors the original supermodel as the "better" model, primarily because the modest improvement on the familiar VinDr-CXR dataset failed to compensate for the diminished generalization observed on the "Other dataset." With only a 12\% improvement on known diseases, it incurred a 33\% decrease in performance on other diseases. Consequently, the supermodel without attention mechanisms proves superior in terms of overall generalization.

The decline in the attention-enhanced supermodel's performance on the "Other Dataset" can be attributed to its focus on the 14 disease-specific features, neglecting the commonalities among different diseases. As evident in Fig. 11, column 6 outperforms column 5 in predicting familiar diseases, yet produces inconsistent results on unseen data. This implies that the model may be emphasizing features associated with the 14 diseases and overlooking the broader aspects indicating "abnormality."

While these models exhibited a higher degree of generalization compared to the specialized models, achieving this required a delicate balance between adaptability and competitiveness with current state-of-the-art models on similar data. To assess their relative standing, a comparative analysis was conducted against top-performing models in the Pneumonia Localization challenge on Kaggle (refer to Table 3). The primary objective of this model was not to surpass the performance of disease-specific state-of-the-art models but to establish a baseline for comparison and demonstrate the model's capacity to segment the disease without being explicitly trained on its unique characteristics. Although the original supermodel falls slightly short of the levels achieved by specialized state-of-the-art models, it approaches remarkably close. This proximity indicates the potential for further enhancements in the final model, ensuring even more robust performance, as observed in CheX-Nomaly.

\subsection{\textbf{CheX-Nomaly}}

\subsubsection{\textbf{Siamese Network}}

The Siamese network functions as a binary classification system utilizing a contrastive loss mechanism to discern the similarity between two chest X-rays (CXRs). Specifically, it aims to determine whether the pair of CXRs shares a similar medical condition (indicating the presence of a disease) or exhibits dissimilarity (where one CXR is healthy). During testing, two variants of the Siamese network were evaluated.

\begin{table}[h] 
\centering
\begin{minipage}{0.45\textwidth}
\centering
\caption{Two Siamese Networks with Varied Output Sizes in Binary Classification}
\label{tab:cnn-unet-siamese10}
\begin{tabular}{|l|l|l|}
\hline
\textbf{} & \textbf{CNN Siamese} & \textbf{U-net Siamese} \\ \hline
\textbf{MAE} & 0.1381 & 0.1334 \\
\textbf{Accuracy} & 0.908 & 0.905 \\ \hline
\end{tabular}
\end{minipage}
\hspace{0.05\textwidth}
\begin{minipage}{0.45\textwidth}
\centering
\caption{Two Siamese Networks with Varied Output Sizes Transferred to final Chex-Net model}
\label{tab:cnn-unet-mean-iou11}
\begin{tabular}{|l|l|l|}
\hline
\textbf{} & \textbf{CNN Siamese} & \textbf{U-net Siamese} \\ \hline
\textbf{MAE} & 0.0739 & 0.0109 \\
\textbf{Mean IOU} & 0.0039 & 0.7487 \\ \hline
\end{tabular}
\end{minipage}
\end{table}

The first variant embraced a Convolutional Neural Network (CNN) architecture combined with a dense layer featuring 128 output units for each image's base model. The second variant preserved the original image size, generating a 512x512x1 feature map using a U-Net-based architecture. Notably, both models exhibited comparable classification performance, as observed in Table 4. However, the true significance of these models surfaced when they were integrated as inputs to the supermodel, as depicted in Table 5.

Evidently, the model's performance with 128 logits notably lagged behind that of the 512x512 feature map model. This performance disparity can likely be attributed to the limited dataset size available for the model to derive a complete mask from. Despite the similar classification performance demonstrated by both models, no further efforts were undertaken to augment the CNN's performance to align with that of the U-Net model. 

\subsubsection{\textbf{CheX-Nomaly Model}}

Fig. 11 provides a comprehensive illustration of the performance of various models across four distinct input images, each representing unique medical conditions:

\begin{enumerate}
    \item \textbf{Pneumonia:} Performance evaluation on an unseen image with an unfamiliar disease from a different dataset.
    \item \textbf{Numerous Local Diseases on One Patient:} Evaluation on an image featuring multiple local diseases within a single patient.
    \item \textbf{One Disease Visible on Patient:} Evaluation on a scan displaying a single abnormality.
    \item \textbf{No Disease:} Performance analysis on a healthy patient's CXR.
\end{enumerate}

Remarkably, CheX-nomaly consistently surpasses all other models across all four scenarios. It adeptly addresses the challenge of generalization that both the supermodels and specialized models encounter, all while maintaining the prediction performance of the original supermodel. Notably, it excels in predicting anomalies within the actual bounding boxes. Additional predictions on unseen diseases can be referenced in Appendix C, while predictions on the VinDr-CXR test set are available in Appendix B.

The results evidently showcase how CheX-nomaly effectively mitigates inherent limitations observed in the other models, thereby preserving high performance on the diseases for which it has been specifically trained.

\subsubsection{\textbf{CheX-Nomaly vs State-of-the-art models}}

\begin{table}[H]
\centering
\caption{Comparison of CheX-Nomaly on the VinDr-CXR Dataset with State-of-the-Art Models}
\label{tab:mean-iou-models12}
\begin{tabular}{|l|l|}
\hline
\textbf{Models} & \textbf{Mean IOU} \\ \hline
CheX-Nomaly & 0.7487 \\
S² & 0.314 \\
SZI & 0.307 \\
scumed & 0.305 \\
fantastic\_hirarin & 0.303 \\
Watercooled & 0.299 \\ \hline
\end{tabular}
\end{table}

\begin{figure*}
    \centering
    \includegraphics[width=0.8\textwidth]{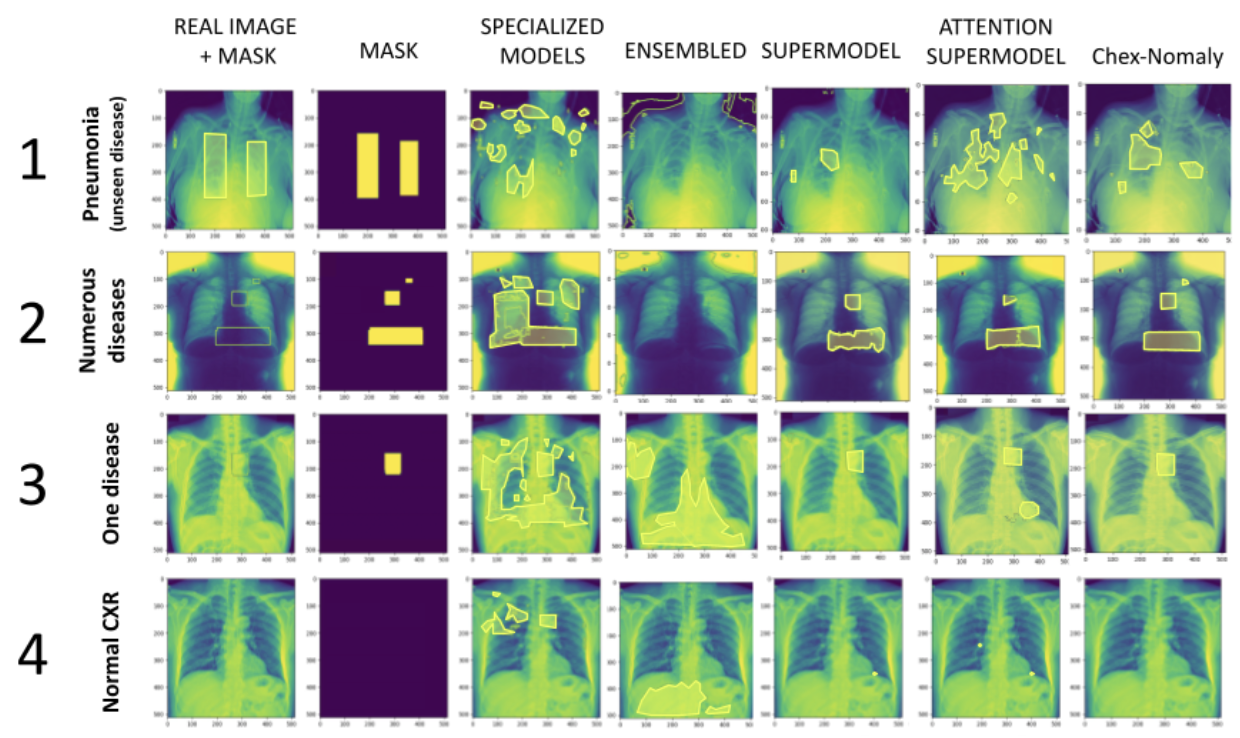} 
    \caption{Visual Comparison of All Models/Methods on Four Test Cases — Pneumonia (Unseen Disease), Numerous Abnormalities, One Disease, and Healthy CXR.}
\end{figure*}

Table 6 offers a comprehensive comparison of CheX-Nomaly's performance on the VinDr-CXR dataset against various state-of-the-art models. In this analysis, CheX-Nomaly's evaluation metric is the Mean Intersection over Union, while the Mean Absolute Precision (mAP) calculated at Intersection over Union (IoU) values exceeding 0.4 is employed for the other models. These other models utilize mask R-CNN and Faster R-CNN object detection methods and hence rely on mAP as their metric of performance. However, it's noteworthy that the U-Net model used in CheX-Nomaly doesn't encompass classification or object detection features.

The employment of IoU as the metric allows for a comparable analysis, addressing the model's focus on "where" rather than "where and what." This difference in the approach impacts the lower mAP value, considering the unique characteristics of the current models that handle both localization and classification of 14 diseases.

Table 3 provides a comparative analysis of CheX-Nomaly's performance on the pneumonia Kaggle dataset alongside state-of-the-art models. Despite not being explicitly trained for pneumonia recognition, CheX-Nomaly demonstrates a competitive Mean IoU of 0.2801. This outcome underscores the model's significant potential for generalization across diverse diagnostic tasks and emphasizes its suitability for a broad range of medical applications.

\begin{figure}[H]
    \centering
    \includegraphics[width=0.48\textwidth]{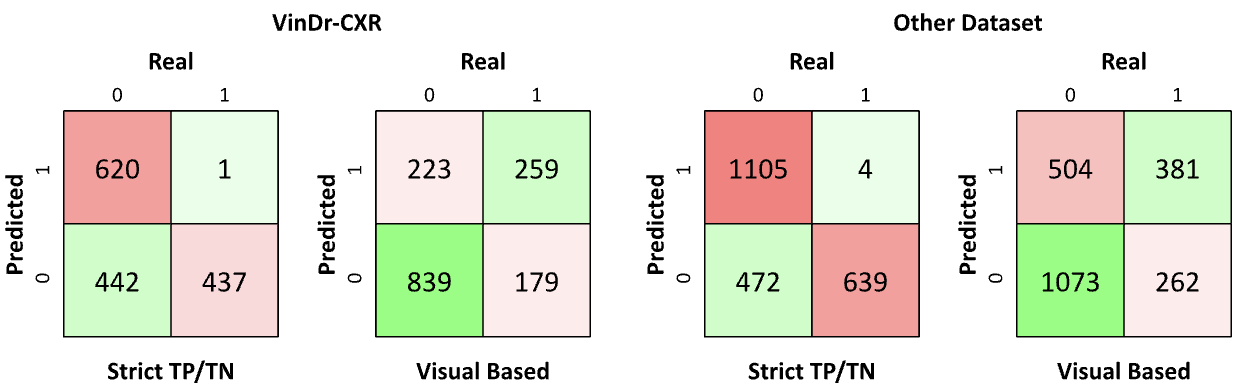} 
    \caption{The confusion matrix above illustrates CheX-nomaly's performance on two datasets, considering strict categorization with perfect true negatives and perfect true positives, as well as a more visually intuitive, less strict categorization.}
\end{figure}

In summary, the performance analysis presented in these tables underscores CheX-Nomaly's proficiency as a versatile and competent model for medical image analysis. Its capacity to deliver superior results, even on datasets featuring diseases for which it hasn't been specifically trained, speaks to its potential for broader clinical applications and establishes its significance as a valuable tool in the realm of medical imaging.

However, it's crucial to acknowledge that CheX-Nomaly, like any model, is not devoid of errors. A comprehensive analysis of these errors is detailed in Appendices B and C.

In evaluating the model's performance on the VinDr-CXR test set, I observed a minimal number of false negatives and notable positive recall. To categorize the predictions, the criteria used involved four categories: True Positives (TP), True Negatives (TN), False Negatives (FN), and False Positives (FP), akin to human interpretation. For instance, some False Positives comprised small, scattered dots of abnormality, easily recognizable as a True Negative to a human observer.

Identifying a True Positive in a segmentation model is typically challenging, as achieving perfect masks for medical problems is inherently difficult. Nonetheless, if the abnormal area corresponds to the radiologist's or human user's expectations, it offers crucial insights into the nature of the disease. The shape need not adhere to a perfect rectangle.

Figure 11 illustrates the model's confusion matrices on both the VinDr-CXR test set and the Other dataset. Additionally, for a more in-depth analysis, histograms in Figure 12 showcase the distribution of error percentages across the entire dataset. The left-skewed curve in both histograms indicates a successful model, with fewer scans showing higher error percentages.

Remarkably, the similarity between the curves on the VinDr-CXR test set and the Other dataset, with slightly more error in the "Other dataset" due to its larger sample size, corroborates the model's success. Both datasets maintain similar ratios of positive to negative data points, enabling meaningful comparisons. This uniformity in the curve shape underscores CheX-Nomaly's effective generalization across various thoracic abnormalities.

\begin{figure}
    \centering
    \includegraphics[width=0.48\textwidth]{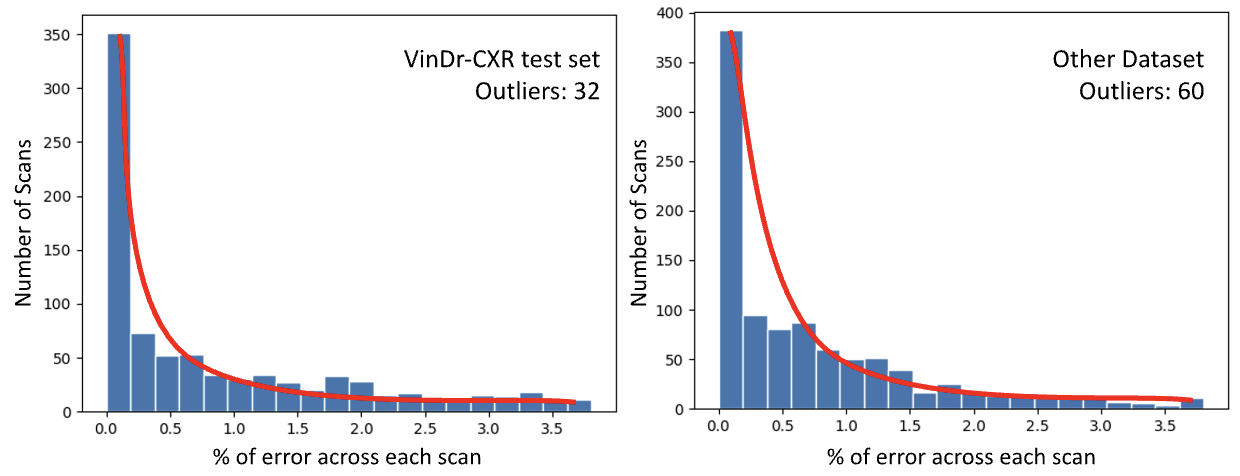} 
    \caption{The histogram shows the percentage of pixels inaccurately classified out of 262144 (512x512) pixels. Cases with more than 4\% error are considered outliers.}
\end{figure}

\section{Discussion}

In the domain of healthcare and artificial intelligence, the scarcity of data presents a pervasive challenge. As a result, AI models undergo extensive clinical trials before their deployment in real-time medical settings. A fundamental goal for all AI models is to achieve generalizability, facilitating their application across a spectrum of healthcare issues, thereby enhancing their utility and effectiveness within the healthcare landscape.

The pursuit of generalized models necessitates the creation of datasets that embody a comprehensive scope. The release of extensive datasets like VinDR-CXR serves this purpose. However, even such expansive datasets have inherent limitations. Models optimized on such datasets may lack familiarity with diseases not encompassed in the dataset. Therefore, the development of computational techniques is essential to guide models away from disease-specific features, encouraging them to adopt a more comprehensive understanding rooted in the appearance of a healthy chest X-ray (CXR). The primary objective is to prompt the model to identify any deviations from this healthy baseline, thereby improving its ability to detect anomalies in CXRs, irrespective of the specific disease targeted.

Further research in this domain might involve:

\begin{itemize}
  \item Development of an Affordable, Efficient, and Real-time Computational Platform
  \item Enhancing Interoperability and Seamless Integration
  \item Exploration of Alternative Contrastive Learning Techniques
\end{itemize}

Integration of expertise from professional clinicians, including doctors and radiologists, can significantly reduce perceptual errors, thus reducing the occurrence of misdiagnoses in thoracic diseases. Additionally, access to private medical datasets, encompassing diverse pathological conditions, holds promise for potential enhancements in model performance and its generalization capabilities.

\section{Conclusion}

Lung and thoracic diseases present a significant challenge, with a notable frequency of misdiagnoses occurring regularly. Focusing on addressing the pervasive issue of perceptual errors is of utmost importance, constituting a significant portion, around 80\%, of diagnostic inaccuracies in chest X-rays. While existing models primarily concentrate on disease classification, accounting for merely 13\% of misdiagnoses, the development of models targeting perceptual errors specifically presents a promising avenue for improvement.

The creation of an effective model to mitigate perceptual errors necessitates a paramount attribute: generalizability. This attribute enables accurate localization of abnormalities across a spectrum of diseases, including those not encountered during model training. CheX-nomaly, an exemplary transfer learning model, leverages a pre-trained Siamese network to discern similarities between different diseases and distinctions between diseases and healthy patients. This unique approach facilitates the acquisition of normal chest X-ray (CXR) features, transcending disease-specific characteristics.

Demonstrating impressive performance metrics, such as a mean absolute error of 0.0109 and a mean intersection over union of 0.7487, models like CheX-nomaly hold the potential to significantly enhance the accuracy of thoracic disease diagnosis.


\newpage

\section{ }
\begin{figure}
    \centering
    \includegraphics[width=0.48\textwidth]{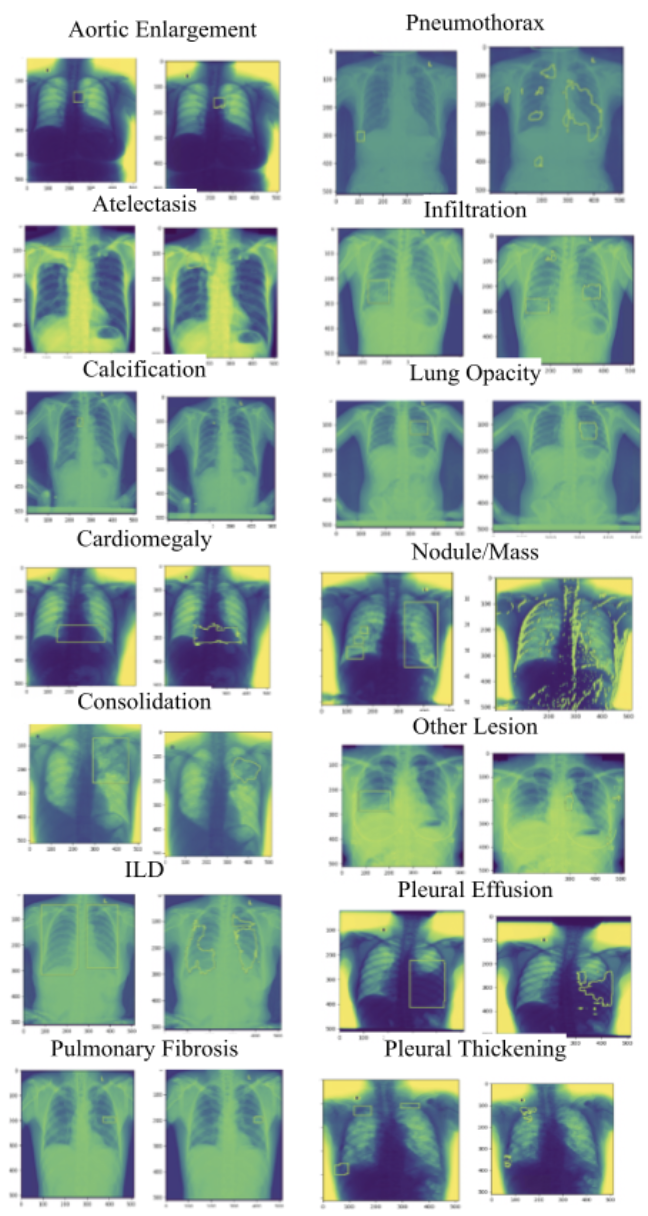} 
    \caption{Appendix A: On the left, the figure illustrates the specialized model's performance on a sample test case, focusing on the corresponding disease. Segmentation outcomes are primarily confined within the bounding box region, demonstrating high performance. This accomplishment was achieved through the implementation of focal cross-entropy, augmentation techniques, and hypertuning. These adaptations allowed the model to transition from predominantly predicting zero values across the entire lung, aiming for heightened accuracy or minimized loss. Notably, only two diseases, aortic enlargement and cardiomegaly, displayed learning patterns in this context, as the presence of fixed masks in specific regions facilitated predictions. After these adaptations, employing focal cross-entropy, augmentation, and hypertuning, the model exhibited a notable shift towards predicting values of one. However, challenges emerged in accurately segmenting certain diseases, such as calcification and pneumothorax, characterized by smaller bounding boxes.}
\end{figure}

\begin{figure}
    \centering
    \includegraphics[width=0.48\textwidth]{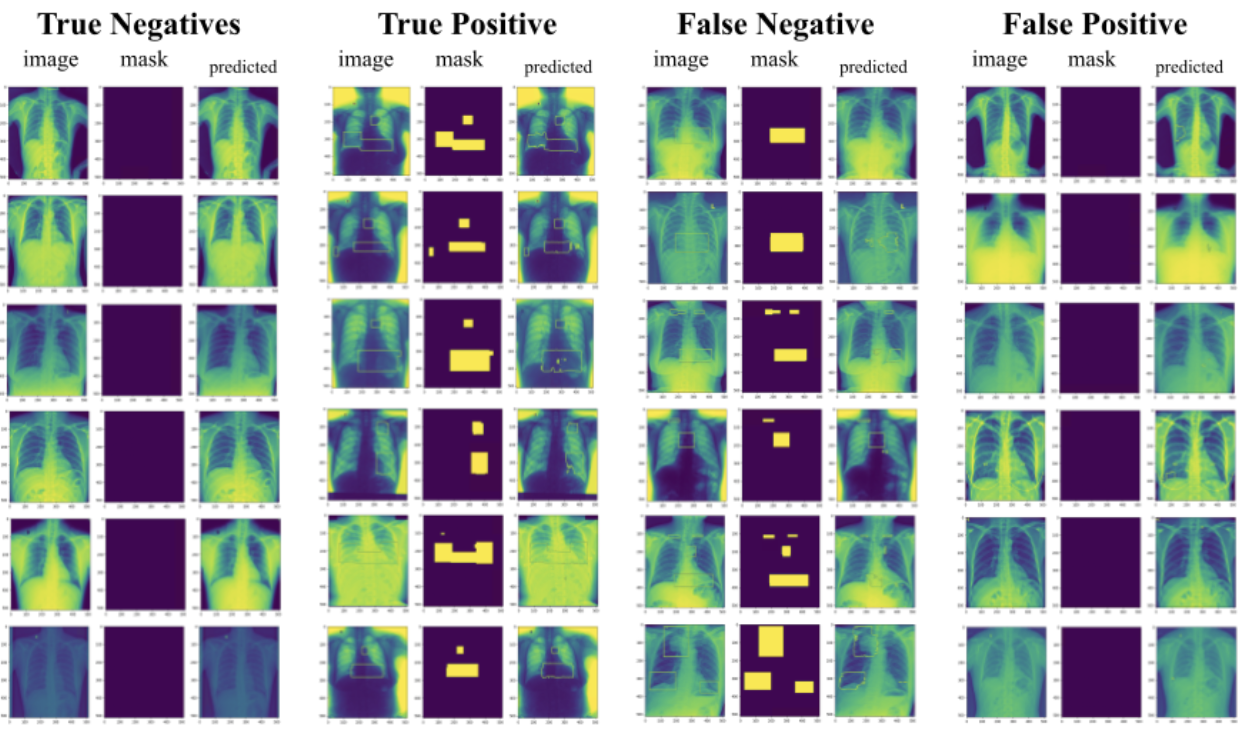} 
    \caption{Appendix B: The images above provide an illustration of CheX-Nomaly's performance across four distinct categories: True Negatives, True Positives, False Negatives, and False Positives. True Negatives and True Positives exemplify images with masks that exhibit near-perfection, reflecting the model's accuracy. True Negatives are when both the mask and prediction have no pixels predicted as 1. Perfect True Positive masks are rare especially due to the irregular abnormality shape of a perfect rectangle hence if the difference between the real mask and prediction was less than 100 pixels out of the 262144 then it would fall under True Positives. On the other hand, False Negatives and False Positives showcase instances of more noticeable errors. False Positives may be comparatively harder to discern visually, given that their predictions are often distributed across the image. In real-world applications, these predictions tend to seamlessly blend into the overall visual output, making it less apparent that no abnormalities were detected. In contrast, False Negatives pose a greater concern, as they can potentially lead to misdiagnosis and other clinical issues. It is worth noting that False Negatives were relatively uncommon in the VinDr-CXR dataset.}
\end{figure}

\begin{figure}
    \centering
    \includegraphics[width=0.48\textwidth]{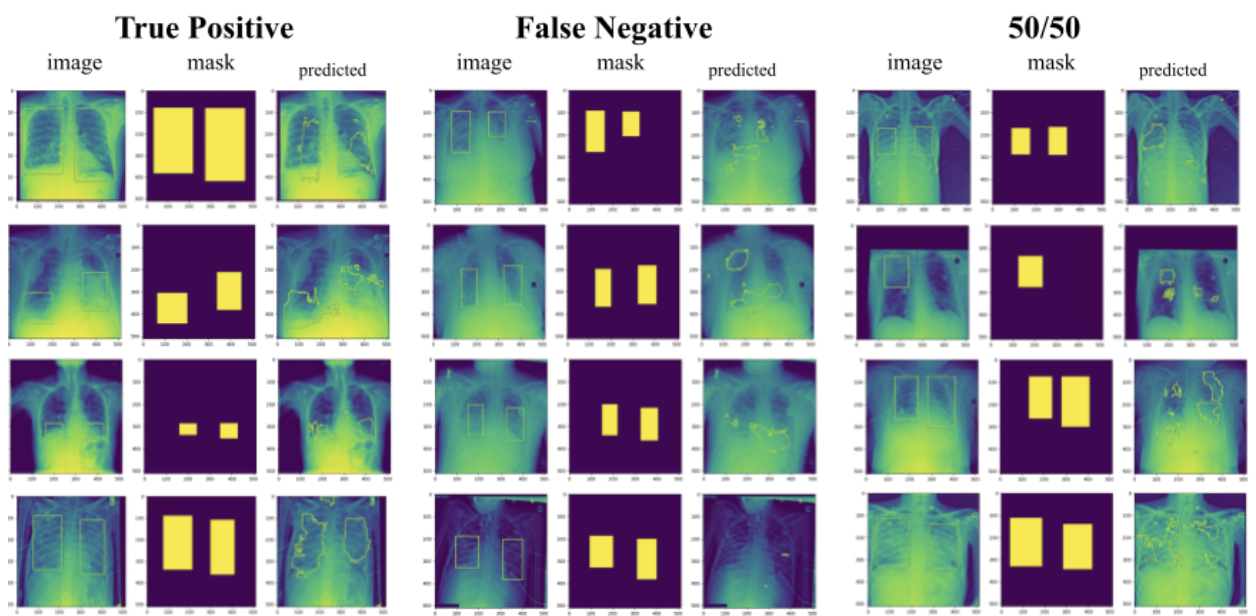} 
    \caption{Appendix C: The images presented above serve as an illustration of CheX-Nomaly's performance on the "Other" dataset, which predominantly comprises images of pneumonia. This dataset was categorized into three distinct groups: True Positives, False Negatives, and 50/50 cases. Notably, the absence of "no disease" images within the dataset precluded their use as negative samples. However, the notion of Negative images, which represent the norm in the context of medical imaging, was consistently prevalent across all patients, making it a well-defined familiar class. The "Other" dataset was meticulously designed to evaluate the model's performance on unfamiliar diseases, particularly focusing on pneumonia, COVID-19, and rib fractures. Within this dataset, True Positive cases displayed visually impeccable masks, with the model adeptly identifying abnormal regions. Conversely, False Negative cases demonstrated a substantial disparity, as they exhibited minimal or no disease identification. These cases might erroneously suggest that the patients are in good health. A subset of cases fell into the 50/50 category, featuring images with multiple bounding boxes. In these instances, the model successfully identified some sections of abnormality, while others remained undetected, reflecting its partial success in detecting the entirety of abnormalities.}
\end{figure}

\end{document}